\documentclass[journal,twoside,web]{ieeecolor}
\usepackage{tmi}
\usepackage{cite}
\usepackage{amsmath,amssymb,amsfonts}
\usepackage{graphicx}
\usepackage{textcomp}
\usepackage{multirow}
\usepackage{multicol}
\usepackage{makecell}
\usepackage[noend]{algpseudocode}
\usepackage{algorithmicx,algorithm}

\def\BibTeX{{\rm B\kern-.05em{\sc i\kern-.025em b}\kern-.08em
    T\kern-.1667em\lower.7ex\hbox{E}\kern-.125emX}}
\begin{document}
\title{Personalized and privacy-preserving federated heterogeneous medical image analysis with PPPML-HMI}

\author{Juexiao~Zhou, Longxi Zhou, Di Wang, \IEEEmembership{Member, IEEE}, Xiaopeng Xu, Haoyang Li, Yuetan Chu, Wenkai Han, Xin~Gao, \IEEEmembership{Member, IEEE}
\thanks{
The research reported in this publication was supported by the King Abdullah University of Science and Technology (KAUST) with grants from the Office of Research Administration (ORA) at KAUST under award numbers FCC/1/1976-44-01, FCC/1/1976-45-01, URF/1/4352-01-01, REI/1/4742-01-01 REI/1/4473-01-01, URF/1/4663-01-01, BAS/1/1689-01-01 and funding from the AI Initiative REI/1/4811-10-01. Corresponding author: Xin Gao. Juexiao Zhou and Longxi Zhou are co-first authors.
}
\thanks{
J. Zhou, L. Zhou, D. Wang, X. Xu, H. Li, Y. Chu, W. Han, and X. Gao are with the Computer Science Program, Computer, Electrical and Mathematical Sciences and Engineering Division, King Abdullah University of Science and Technology (KAUST), Thuwal 23955-6900, Kingdom of Saudi Arabia and Computational Bioscience Research Center, King Abdullah University of Science and Technology, Thuwal 23955-6900, Kingdom of Saudi Arabia. (e-mail: xin.gao@kaust.edu.sa).
}
}

\maketitle

\begin{abstract}
Heterogeneous data is endemic due to the use of diverse models and settings of devices by hospitals in the field of medical imaging. However, there are few open-source frameworks for federated heterogeneous medical image analysis with personalization and privacy protection simultaneously without the demand to modify the existing model structures or to share any private data. In this paper, we proposed PPPML-HMI, an open-source learning paradigm for personalized and privacy-preserving federated heterogeneous medical image analysis. To our best knowledge, personalization and privacy protection were achieved simultaneously for
the first time under the federated scenario by integrating the PerFedAvg algorithm and designing our novel cyclic secure aggregation with the homomorphic encryption algorithm. To show the utility of PPPML-HMI, we applied it to a simulated classification task namely the classification of healthy people and patients from the RAD-ChestCT Dataset, and one real-world segmentation task namely the segmentation of lung infections from COVID-19 CT scans. For the real-world task, PPPML-HMI achieved $\sim$5\% higher Dice score on average compared to conventional FL under the heterogeneous scenario. Meanwhile, we applied the improved deep leakage from gradients to simulate adversarial attacks and showed the solid privacy-preserving capability of PPPML-HMI. By applying PPPML-HMI to both tasks with different neural networks, a varied number of users, and sample sizes, we further demonstrated the strong robustness of PPPML-HMI.
\end{abstract}


\section{Introduction}\label{sec:introduction}
\IEEEPARstart{D}{ata-hungry} artificial intelligence (AI), including various machine learning (ML) and deep learning (DL) methods\cite{lecun2015deep}, is increasingly being applied to solve miscellaneous tasks in medical image analysis (MIA) and has led to disruptive innovations in pathology, radiology, and other fields \cite{greenspan2016guest, ting2018ai, lundervold2019overview, rieke2020future, zhou2021review, aggarwal2021diagnostic, zhou2022interpretable}. However, since modern DL models typically have millions or even more parameters\cite{rasley2020deepspeed}, a mass of curated data is usually required to train such data-hungry models to achieve clinical-grade performance \cite{sun2017revisiting, aggarwal2018neural, wang2019deep}. In contrast, even with modern advanced data science, generating a huge amount of data to train models independently is still challenging for most hospitals and clinics. Therefore, seeking the cooperation of institutions to jointly generate data and train a joint model becomes an ideal solution \cite{mammen2021federated}. In centralized training, the server needs to collect data from all collaborators. Then the model will be trained on the server. Nevertheless, such a strategy leads to concerns related to data security and privacy. For example, training an AI-based lung infection detector \cite{shan2020lung, zhou2020rapid, serte2021deep, saood2021covid} required a large amount of high-quality computerized tomography (CT) scans and human-labelled metadata. However, such data is difficult to be obtained and shared in reality because health data is usually highly sensitive and tightly regulated \cite{van2014systematic, rieke2020future}. Hence, federated learning (FL) \cite{konevcny2016federated, mcmahan2017communication, yang2019federated, li2020federated} was proposed as a learning paradigm that aims to address data governance and privacy issues by collaboratively training models without the need to share the data itself.

$\bold{Federated\ learning}$ In FL, it is assumed that a set of $n$ ($n\ge2$) users are connected to a server, where each user can only access its own data \cite{konevcny2016federated}. Upon that, the users' goal is to acquire a model that captures the features of all users' data without sharing their local data with any other user or server. Though each user can solely train a model with its own data, the independently trained model of each user may not generalize well to other users' data or new samples. Thus the following FL procedure was proposed to learn a more generalized server model. Firstly, all users will receive a copy of the current server model and update the local model using its own data. After that, users send the updated model to the server. Finally, the server aggregates received local models to update the server model for the next broadcasting. This process continues until a generalized server model could be generated \cite{mcmahan2017communication, mohri2019agnostic}. To be more specific, McMahan et al. \cite{mcmahan2017communication} proposed the federated averaging (FedAvg) algorithm, which is the most famous aggregation method in the community, to aggregate local models collected by the server. 

However, previous studies showed that the FedAvg algorithm might not converge or could be slowed down when local models drift significantly from each other due to the heterogeneity of local non-independent and identically distributed (non-IID) data \cite{karimireddy2019scaffold}. Therefore, in the presence of heterogeneity, the server model trained by FL may not generalize well to each user's data \cite{jiang2019improving}, which is a significant obstacle to applying FL in practice. Taking the infection segmentation on CT scans as an example, different hospitals may have diverse CT scanners and scanning settings, thus the CT scans will have heterogeneity. With that, the server model trained with FL would be unable to achieve clinical-grade performance on each user's data.

$\bold{Personalized\ federated\ learning}$ To apply the FL paradigm with the heterogeneous data as in the case of CT diagnosis, personalized federated learning (PFL) was devised as an enhanced version of FL \cite{hanzely2020lower, fallah2020personalized, t2020personalized, mansour2020three, sun2021pain, li2021ditto}. To address personalization in FL, a two-step approach namely `FL training + local adaptation' was regarded as the most commonly acknowledged strategy by the FL community \cite{mansour2020three, kairouz2021advances}. With this strategy, the server model is firstly trained using FL on heterogeneous users' CT data. Unexpectedly, the server model may perform poorly on each user's data due to data heterogeneity. Therefore, a few additional training steps are required to adapt this server model locally and realize the personalization. Depending on the specific strategies used in training, different personalized variants of the FedAvg algorithm were proposed, such as pFedMe \cite{t2020personalized}, Per-FedAvg \cite{fallah2020personalized} and APFL \cite{deng2020adaptive}. However, all the aforementioned methods were theoretical research, and a little applied attempt has been conducted, especially in the medical analysis field. In addition to optimizing the training strategies for heterogeneous data, FedAVG+Share \cite{zhao2018federated} improves performance on non-IID data by sharing a small amount of data among users. However, the strategy could not be adopted when the user's data is required to be strictly private. FedReplay \cite{qu2022handling} needs to train a universal and auxiliary encoder network, which encodes each user's data into latent variables that will be used to train the server model for classification. Therefore, given an existing neural network model, e.g. a segmentation model, it needs to be disassembled and restructured to work with FedReplay in FL. Thus FedReplay could not be simply and directly combined with the existing models and also might not be easily used by new users as a closed-source method, which might limit its wide usage. As the latest work, FedPerGNN \cite{wu2022federated} was specially designed for graph neural networks and thus was limited to the application of graph data, which is usually different from medical imaging data.

$\bold{Privacy\ in\ FL\ and\ PFL}$ Privacy is a hot and significant topic in the age of medical big data \cite{price2019privacy}. Nevertheless, previous studies showed that FL is still vulnerable to attacks, such as data poisoning attack \cite{tolpegin2020data}, membership inference attack \cite{rahman2018membership, lu2020sharing, salem2018ml}, source inference attack (SIA) \cite{hu2021source}, attribute reconstruction attack \cite{lyu2021novel}, and inversion attack\cite{geiping2020inverting, zhao2020idlg, yin2021see, hatamizadeh2022gradient}, thus compromising data privacy. 

Similar to FL, PFL is also facing the threat of privacy attacks. Diverse strategies could be used to protect data privacy with FL and PFL. As the latest research, differential privacy (DP) \cite{dwork2008differential} was used to add noise to the gradient transmitted in PFL \cite{hu2020personalized, wu2022federated} as same as in FL \cite{zhou2022ppml} to protect the privacy of users' data. However, DP adopts the mechanism of adding noise to enhance privacy protection while sacrificing model accuracy \cite{bagdasaryan2019differential}, resulting in difficulty to achieve clinical-grade performance in practice. Besides, cryptographic approaches, including secure aggregation (SA) \cite{shamir1979share, bonawitz2017practical}, homomorphic encryption (HE) \cite{gentry2009fully}, multi-party computation (MPC) \cite{Goldreich} and etc., realize privacy preservation by sacrificing time and space without affecting the accuracy much. Among these techniques, MPC requires the involvement of multiple servers, which is different from the case with only one server. Conventional SA requires heavy communications between users and the server, which will cause overhead for users with limited resources, while the latest decentralized version of SA \cite{sandholm2021safe} transferred the process of aggregation from the server to users, reduced the need for high communication between all users and the server, and further strengthened the privacy protection as the server is usually an un-trusted third party. However, the protocol used in \cite{sandholm2021safe} required a three-step process as `decryption-summation-encryption' to aggregate the local gradients into the transmitted gradients in the loop, which could be further simplified into a two-step process as `encryption-summation' with the help of HE as in PPPML-HMI. 

To strengthen privacy protection in FL, several open-source methods, such as FATE\cite{fate}, PySyft\cite{Ziller2021}, and NVFlare\cite{roth2022nvidia}, have already been developed to secure gradients during training using techniques like SA and HE. However, these methods are mainly based on the conventional FL and thus could not address the challenges with heterogeneous data while still allowing privacy protection, because personalization and privacy protection are still not addressed simultaneously under the federated scenario.

\begin{figure}[htb]
\centering
\includegraphics[width=0.5\textwidth]{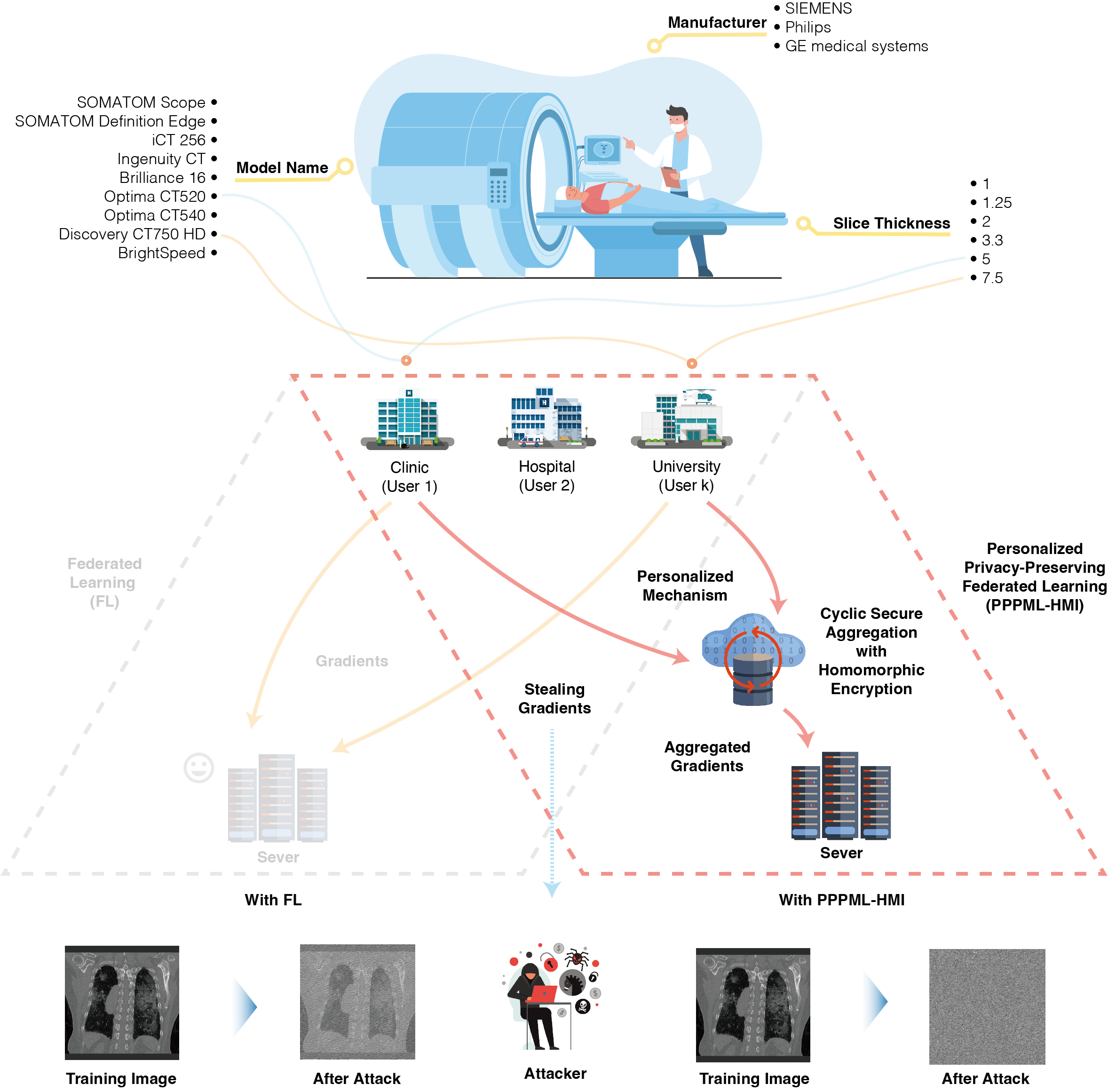}
\caption{Scheme of PPPML-HMI. In our real-world case, hospitals use devices from different manufacturers with various models and settings for the detection of lung infection by COVID-19. The use of diverse devices generates data with inherent differences, namely heterogeneous data. With FL, the goal is to jointly train a consensus model with the data from each hospital without sharing the data itself. With homogeneous data across hospitals, FL could efficiently train a server model that works well for all hospitals. However, when hospitals have heterogeneous data, the server model trained by FL could not perform well when applied to each hospital. Thus, PPPML-HMI allows models to adapt to heterogeneous data. To strengthen the privacy protection of PPPML-HMI, we designed the cyclic secure aggregation with homomorphic encryption.}
\label{fig1}
\end{figure}

$\bold{Heterogeneity\ and\ privacy\ in\ MIA}$ As a representative of many medical tasks that would strongly benefit from personalization and privacy protection, especially in medical imaging, the accurate detection and segmentation of lung infection caused by the severe acute respiratory syndrome coronavirus 2 (SARS-CoV-2, COVID-19) have been such an important task since 2020 \cite{fauci2020covid}. To diagnose lung diseases, imaging is the major source of data and the most commonly used imaging technologies are X-rays and CT scans \cite{franquet2001imaging}. Meanwhile, DL has been widely applied in developing the computer-aided diagnosis (CAD) systems for COVID-19 \cite{zhou2020rapid, wang2020fully, jamshidi2020artificial, shorten2021deep, ismael2021deep, zhou2022interpretable, subramanian2022review}. Most aforementioned works require centralized training, where the research institution acts as a coordinator/server to collect raw CT scans from users like hospitals to train a model centrally. Several problems exist in this process. Firstly, all users need to strictly trust each other and the server in order to share the raw CT scans, which might limit the number of users and available data involved, leading to insufficient training data. Secondly, users have to deliver the private raw CT scans to the server, leading to potential privacy breaches. Therefore, a proper FL approach is necessary to allow users to keep their data private, thus more data providers could participate in the co-training of the model and more diverse data could be used to train a model with better generalization power. Though FL has been widely applied in tasks related to COVID-19 in the latest research \cite{dayan2021federated, durga2022fled, samuel2022iomt, florescu2022federated, li2022integrated, wibawa2022homomorphic}, the heterogeneous data and privacy breaches are still problems as there is currently no such open-source solution for personalized and privacy-preserving federated heterogeneous medical image analysis, especially for the heterogeneous COVID-19 CT analysis.

Here, we proposed PPPML-HMI, a novel open-source, robust, user-friendly and plug-and-play method for personalized and privacy-preserving federated heterogeneous medical image analysis ($\bold{Figure}\ref{fig1}$). PPPML-HMI specifically targets the scenario where no raw data should be shared with any third party, no structural modification should be conducted to existing DL models, and in a context of heterogeneous data. To our best knowledge, personalization and privacy protection were achieved simultaneously for the first time under the federated scenario by integrating the PerFedAvg algorithm and designing the novel cyclic secure aggregation algorithm with homomorphic encryption (CSAHE). To demonstrate the utility of PPPML-HMI, we applied it to a simulated classification task namely the classification of healthy people and patients from the RAD-ChestCT Dataset \cite{draelos2021machine}, and one real-world segmentation task namely the segmentation of lung infections from COVID-19 CT scans by extending our previous method for the task \cite{zhou2020rapid, zhou2022interpretable}, which was also a general method for segmentation of lung, tracheal, vascular, etc. By applying PPPML-HMI to both tasks with different neural networks, a varied number of users, and sample sizes, we further demonstrated the robustness of PPPML-HMI. Finally, we also applied the improved deep leakage from gradients to simulate adversarial attacks and showed the strong privacy-preserving capability of PPPML-HMI.

\section{Methods}
\subsection{Design of PPPML-HMI}\label{designPPPML_HMI}
PPPML-HMI is built up with two major modules as a training framework: the PFL ($\bold{Section\ \ref{method_pfl}}$) and the CSAHE modules ($\bold{Figure\ \ref{fig1}}$, $\bold{Algorithm\ \ref{algo_PPPML_HMI}}$ and $\bold{Section\ \ref{method_csahe}}$). During each round of the global training, the server broadcasts the server model to each user for initialization. Then, each user trains the local model using the local private data. After finishing the local training, each user calculates the gradient between the local model and the server model. To avoid sending users' gradients directly to the server, which could lead to potential privacy leakage, PPPML-HMI transfers the gradient aggregation process that is originally performed on the server to a loop composed of all users through the CSAHE mechanism in a decentralized manner. At the end of each global training, the CSAHE mechanism is executed. A user in the loop will be randomly selected as the initiator, who will protect its own gradient by summing a random mask as noise to the gradient and encrypting the noised gradient with HE, and will transmit the noised gradient into the loop for further aggregation. The noise in the aggregated gradients is kept till the end of the execution of the CSAHE mechanism. PPPML-HMI achieves decentralized secure gradient aggregation with homomorphic encryption ($\bold{Section\ \ref{method_csahe}}$), thus each user could confidently aggregate their own gradients to the transmitted gradient without worrying about privacy issues. The code of PPPML-HMI is publicly available at https://github.com/JoshuaChou2018/PPPML-HMI.

\subsection{Dataset processing}

For the classification task, we simulated and constructed our heterogeneous data from the RAD-ChestCT Dataset \cite{draelos2021machine, draelos_rachel_lea_2020_6406114}, which includes 35,747 chest CT scans from 19,661 adult patients. For the segmentation task, we collected 180 anonymized CT scans generated by diverse CT scanners and scanning parameters from five hospitals labelled as $A \sim E$. All patients were confirmed to be COVID-19 positive by either the nucleic acid test or antibody test. For the classification task, we used the publicly available RAD-ChestCT Dataset \cite{draelos2021machine, draelos_rachel_lea_2020_6406114}. For the segmentation task, the data from partner hospitals are available upon request.

To perform the segmentation of lung infections from the 3D CT scans, we need to find a mapping $\mathbb{F}: \mathbb{R}^{H \times W \times S} \mapsto \{0,1\}^{H \times W \times S}$, where $H \times W$ is the height and width of each 2D CT image and $S$ is the number of images. Since the data generated by different CT scanners owned various volume sizes, spatial normalization was adopted to re-scale raw CT data into a machine-agnostic standard space with fixed shape $(512\times512\times512)$. As in \cite{zhou2020rapid}, we decomposed the 3D segmentation of each 3D CT scan into three 2D segmentation problems along the x-y, y-z, and x-z views (axial, sagittal, and coronal). The training along each plane was performed independently. All prediction and visualization of samples were performed with models trained with data excluding the corresponding sample itself.

\begin{figure}[htb]
\centering
\includegraphics[width=0.5\textwidth]{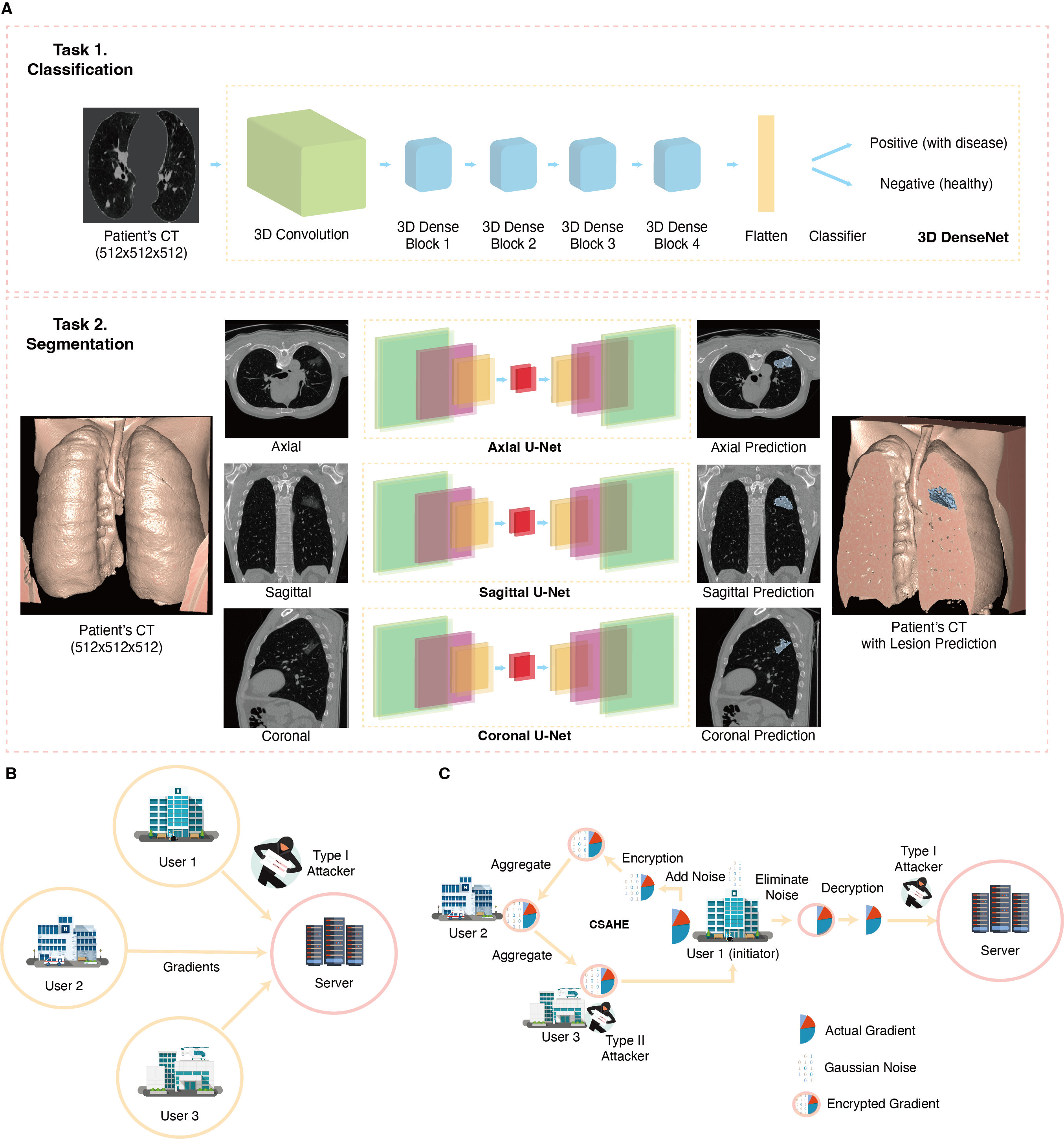}
\caption{A) Illustration of two tasks.  We applied PPPML-HMI to the classification of healthy people and patients with 3D DenseNet on the RAD-ChestCT Dataset, and the segmentation of the lung infections of COVID-19 with a 2.5D U-Net method \cite{zhou2020rapid, zhou2022interpretable}. Illustration of the communication network and attackers of FL (B) and PPPML-HMI (C). Two types of attackers exist in our setting: 1) Attackers who can intercept messages sent from any users to the server or between users (type I), and 2) Honest-but-curious attackers who are part of the users of PPPML-HMI (type II).}
\label{fig2}
\end{figure}

\subsection{Neural network for the segmentation of lung infections}

For the classification task, we adopted the 3D DenseNet \cite{3ddensenet} as the backbone DL model.

For each segmentation task along the three views, we trained an independent U-Net that took five adjacent images with dimension: $\mathbb{R}^{5\times512\times512}$ as inputs, and output the probability map of infection regions for the central image with dimension: $\mathbb{R}^{512\times512}$. The U-Net for 2D segmentation consisted of four encoding layers, one bottleneck layer, and four decoding layers as shown in $\bold{Figure \ref{fig2}A}$. 

Given a 3D CT scan, we applied three 2D U-Net models and generated three segmentation results $p_{xy}, p_{yz}, p_{xz}$ along the x-y, y-z, and x-z views. The final segmentation result in 3D space was calculated by summing up three intermediate predictions followed by taking a threshold of 2 as $p_{final} = (p_{xy} + p_{yz} + p_{xz}) \geq 2$.

\subsection{Personalized federated learning}\label{method_pfl}

To accomplish personalized FL, the personalized FedAvg (Per-FedAvg) \cite{fallah2020personalized} algorithm was adopted to acquire the optimal initial model (meta-model) as the server model, which could be easily adapted to the local heterogeneous data by performing just a few steps of gradient descent. Per-FedAvg was inspired by the fundamental idea of the Model-Agnostic Meta-Learning (MAML) framework \cite{finn2017model}. Given a set of tasks from different underlying distributions, instead of finding the model that generalizes on all tasks as FL, MAML tends to find a meta-model that could perform better in different tasks after a few steps of local gradient descent. 

In FL, the goal of optimization is $\min_{w \in R^d} f(w) = \frac{1}{n} \sum_{i=1}^nf_i(w)$, where $f_i$ is the loss function to user $u_i$. With the concept of MAML, the goal of the optimization becomes finding a good initialization as $\min_{w \in R^d} F(w) = \frac{1}{n} \sum_{i=1}^nf_i(w-\alpha \triangledown f_i(w))$, where $\alpha$ ($\alpha \ge 0$) is the step size.

As shown in $\bold{Algorithm \ref{algo_PPPML_HMI}}$, at each epoch $k$, the server will broadcast the server model to all users. Then, all users will train their local model with $\tau$ local epochs. After $\tau$ local epochs, a list of $\{w_{k+1,t}^i\}_{t=0}^\tau$ will be generated with respect to the user $u_i$, where $w_{k+1,0}^i = w_k$, $w_{k+1,t}^i = w_{k+1,t-1}^i - \beta \widetilde\triangledown F_i(w_{k+1,t-1}^i)$, $\beta$ is the local learning rate and $\widetilde\triangledown F_i(w_{k+1,t-1}^i)$ is an estimate of $\triangledown F_i(w_{k+1,t-1}^i)$.

\begin{algorithm}
    \caption{PPPML-HMI}
    \scriptsize
    \label{algo_PPPML_HMI}
    \begin{algorithmic}[1]
    \Require $K$ is the number of global epochs;  $\tau$ is the number of local epochs; $N$ is the number of users; $\gamma$ is the number of final adaptation epochs; $w_k$ is the parameters of the server model at global epoch $k$; $w_{k,t}^i$ is the parameters of the model of user $u_i$ at local epoch $t$ and global epoch $k$; $\gamma$ is the number of epochs for the final adaptation.
    \Procedure{System Starts}{} 
        \For {$k=0,\dots,K-1$ (Global epoch)}
            \State Server sends $w_k$ to all users;
            \For {each user $u_i$ with $i=0,..,N-1$}
            	\State Set $w_{k+1,0}^i=w_k$
                \For{$t=0,...,\tau-1$ (Local epoch)}
                	\State Compute the stochastic gradient $\tilde{\bigtriangledown}f_i(w_{k+1,t-1}^i,D^i)$ using dataset $D^i$
                    \State Set $\tilde{w}_{k+1,t}^i=w_{k+1,t-1}^i-\alpha \tilde{\bigtriangledown}f_i(w_{k+1,t-1}^i,D^i)$
                    \State Set $w_{k+1,t}^i=w_{k+1,t-1}^i-\beta (I-\alpha \tilde{\bigtriangledown}^2f_i(w_{k+1,t-1}^i,D^{''i}))\tilde{\bigtriangledown}f_i(\tilde{w}_{k+1,t}^i,D^{'i})$
                \EndFor
                \State Each user $u_i$ with $i=0,..,N-1$ calculates the gradient compared to the server $\Delta w_{k+1,\tau}^i = (w_{k+1,\tau}^i-w_k)$
                \State Execute cyclic secure aggregation with $\bold{CSAHE}$($u_0, ..., u_{N-1}$)
                \State Initiator $u_I$ sends securely aggregated update $\Delta w_{k+1,\tau}^{CSA} $ back to the server 
            \EndFor
            \State Server updates its model by averaging over the received gradients $w_{k+1}=w_k + \frac{1}{N} (\Delta w_{k+1,\tau}^{CSA})$
        \EndFor
        \State Server broadcasts the meta-model to all users
        \For {each user $u_i$ with $i=0,..,N-1$}
            \For{$t=0,...,\gamma-1$ (Final local adaptation)}
                \State User $u_i$ adapts the meta-model with local private data
            \EndFor
        \EndFor
    \EndProcedure
    \\
    \Function{CSAHE}{$u_0,...,u_{N-1}$}
        \State Organize all users in a loop. Randomly select a user $u_I$ with $I\in\{0,...,N-1\}$ as the initiator, generate a random mask $\bold{R}$ with the same shape as the gradient $\Delta w_{k+1,\tau}^I$ hold by $u_I$ from Gaussian Noise with a large $\sigma$.
        \State User $u_r$ adds the $\bold{R}$ to the gradient as $\Delta w_{k+1,\tau}^{CSA} = \Delta w_{k+1,\tau}^I + \bold{R}$ and encrypt it with homomorphic encryption (HE) as $HE(\Delta w_{k+1,\tau}^{CSA})$
        \For {each user $u_j$ in the circular chain, until $j = I$}
            \State $HE(\Delta w_{k+1,\tau}^{CSA}) = HE(\Delta w_{k+1,\tau}^{CSA}) + HE(\Delta w_{k+1,\tau}^j)$
            \State User $u_j$ transfers the $HE(\Delta w_{k+1,\tau}^{CSA})$ to the next user in the loop
        \EndFor
        \State The initiator $u_I$ removes the random mask from the final aggregated gradient with $HE(\Delta w_{k+1,\tau}^{CSA}) = HE(\Delta w_{k+1,\tau}^{CSA}) - HE(\bold{R})$ and decrypts the encrypted aggregated gradient $HE(\Delta w_{k+1,\tau}^{CSA})$ into $\Delta w_{k+1,\tau}^{CSA}$.
        \State Return the securely aggregated gradient $\Delta w_{k+1,\tau}^{CSA}$
    \EndFunction
\end{algorithmic}
\end{algorithm}

\begin{algorithm}
\caption{iDLG}
\label{algoiDLG}
\scriptsize
\begin{algorithmic}[1]
\Require Differentiable model $\mathbf{M}$, model parameters $\mathbf{W}$, private training data and labels $(\mathbf{x},\mathbf{c})$, gradients produced by the private data $\nabla \mathbf{{W}}$ ,dummy data and labels $(\mathbf{x}',\mathbf{c'})$, number of iterations $N$, learning rate $\eta$ and loss function $l$
\State Extract the target ground-truth label to initialize the dummy label $\mathbf{c'}$
\State Initialize the dummy data $\mathbf{x'} \leftarrow \mathcal{N}(0,1)$
\For{i $\leftarrow$ 1 to N}
    \State Calculate the dummy gradients: $\nabla \mathbf{W} \leftarrow \partial l(\mathbf{M}(\mathbf{x'},\mathbf{W}),c')/\partial \mathbf{W}$
    \State Calculate the loss: $L_G=\|\nabla \mathbf{W}'-\nabla \mathbf{W}\|^2_{F}$
    \State Update the dummy data: $\mathbf{x'}\leftarrow\mathbf{x'}-\eta \nabla_{\mathbf{x'}}L_G$
\EndFor
\end{algorithmic}
\end{algorithm}
\bigskip

\subsection{Cyclic secure aggregation with homomorphic encryption}\label{method_csahe}

We designed the cyclic secure aggregation with homomorphic encryption (CSAHE) algorithm to transfer the secure aggregation from the server to a loop composed of all users in a decentralized manner with a two-step process as `encryption-summation' for all non-initiator users. To protect the gradients transmitted in CSAHE, we encrypted all gradients with homomorphic encryption (MHE) based on the Cheon-Kim-Kim-Song (CKKS) cryptographic scheme \cite{cheon2017homomorphic} that provides approximate arithmetic over vectors of complex numbers and performed the aggregation homomorphically with TenSEAL \cite{benaissa2021tenseal}, which is a python library for performing homomorphic encryption operations on tensors, built on top of Microsoft SEAL.

We integrated the CSAHE into PPPML-HMI to protect the gradients of users. As shown in $\bold{Algorithm \ref{algo_PPPML_HMI}}$, in each epoch, all users form a loop and an initiator is selected randomly from all users, while the remaining users are called non-initiator users. The initiator generates a random mask using a Gaussian distribution with a self-defined large $\sigma$. After that, the random mask will be summed to the initiator's gradient to protect its actual value. Then, the noised gradient will be homomorphically encrypted and transmitted to the next user in the loop, who can also safely aggregate its gradient to this transmitted gradient homomorphically, and transmit the newly aggregated gradient to the next user. The aforementioned process keeps working till the aggregated gradient is transmitted back to the initiator. Then, the initiator eliminates the random mask from the aggregated gradient and decrypts it to recover the actual value. Finally, the initiator sends the aggregated gradient to the server for updating the server model. 

With CSAHE, each user except the initiator only needs to interact with two users located before and after in the loop and does not need to have any interaction with the server. The secure aggregation in PPPML-HMI is conducted in a decentralized manner, which is different from the conventional SA that happens at the server. HE in CSAHE allows all non-initiator users to aggregate their gradients homomorphically without the need to decrypt. As shown in $\bold{Algorithm \ref{algo_PPPML_HMI}}$, once the $w_{k+1,\tau}^i$ of each user $u_i$ is calculated, user $u_i$ will calculate the difference between the $w_{k+1,\tau}^i$ and the server model $w_k$ as $\triangle w_{k+1,\tau}^i = w_{k+1,\tau}^i - w_k$. Then, all $\triangle w_{k+1,\tau}^i$ with $i=0,...,N-1$ will be securely aggregated through the CSAHE algorithm. Finally, the server will collect the securely aggregated gradient $\Delta w_{k+1,\tau}^{CSA}$ from the initiator for updating the server model with $w_{k+1} = w_k + \frac{1}{N}\Delta w_{k+1,\tau}^{CSA}$.

The public key encryption scheme is used by CSAHE, where a public key and a secret key are generated ($\bold{Figure \ref{fig2}C}$). The public key is shared by all users to allow the encryption of gradients before the homomorphic aggregation and the private key is held only by the initiator to allow the decryption of the aggregated gradient before sending it to the server. Although a non-initiator cannot obtain the private key by default, we still need to discuss the vulnerability of CSAHE in practice where the private key could be leaked, thus a non-initiator could also decrypt the homomorphically encrypted gradient when the attack happens.

\subsection{iDLG reconstruction attack}\label{method_idlg}
Previous research has shown that the gradients transmitted from the user to the server in FL may still compromise data privacy\cite{geiping2020inverting, zhao2020idlg, yin2021see}. Among those studies, the improved deep leakage from gradient (iDLG) is a state-of-the-art approach to obtain private training data from the gradients transmitted between users and the server as shown in $\bold{Algorithm \ref{algoiDLG}}$.

\subsection{Performance evaluation}

To evaluate the performance on the classification task, we used the accuracy as defined: $Accuracy = \frac{TP+TN}{TP+TN+FP+FN}$, where TP, TN, FP, and FN stand for true positive, true negative, false positive and false negative.

To evaluate the performance of lung infection segmentation with different methods, we used the Dice score and the recall as defined: $Dice = \frac{2|Y \cap Y'|}{|Y|+|Y'|}$, $Recall = \frac{|Y \cap Y'|}{|Y|}$, where $Y$ is the actual infection region annotated by radiologists, $Y'$ is the predicted infection region, and $|Y|$ represents the cardinality of $Y$. 

\subsection{Hyper-parameter selection and training settings}

To ensure a fair comparison, we tested the number of global epochs from $\{10, 20, 50, 100\}$ and the number of local epochs from $\{1, 5, 10, 20\}$. From our pre-experiments, we noticed that the number of global epochs $K=20$ and the number of local epochs $\tau=10$ enabled the model to converge and provided a good trade-off between the computation time and model performance. The batch size and learning rate were set to be 64 and $10^{-4}$ respectively. For CSAHE, the random noise was generated from a Gaussian distribution with mean $=0$ and a randomly chosen large standard deviation $(>100)$ to avoid the potential inversion attack. During the training, 32 workers were used for data loading and processing on one machine with 120GB RAM and one NVIDIA V100 GPU. Five-fold cross-validation was adopted for data splitting and model training.

\section{Results}

We applied PPPML-HMI to a simulated heterogeneous dataset from the RAD-ChestCT dataset, where users' data were grouped according to the slice thickness from {2 mm, 5 mm, 10 mm}, to train a classification model and show the robustness of PPPML-HMI when varying the number of users and the sample sizes. We also applied PPPML-HMI to a real-world case, where heterogeneous data were generated by five hospitals with different CT scanners, to train a segmentation model for COVID-19 lung infections. To demonstrate the effectiveness of PPPML-HMI, we compared three methods, including training independently using only each user's own data, training in a centralized training manner using complete data, and training with one of the most classical and famous algorithms in FL namely FedAvg \cite{mcmahan2017communication}. With both tasks together with different neural networks, the number of users, and sample sizes, we further gave evidence of the robustness and robustness of PPPML-HMI to these parameters. Meanwhile, we applied the improved deep leakage from gradients to simulate adversarial attacks on the segmentation task and showed the strong privacy-preserving capability of PPPML-HMI.

\begin{table}[!htb]
    \caption{Description of the RAD-ChestCT dataset and averaged predicted accuracy of methods.}
    \label{table_rad}
    \scriptsize
    \centering
    \begin{tabular}{c c c c c c c }
 \hline
 {\makecell{Split\\ID}}  & {\makecell{User\\ID}} & {\makecell{Slice\\thickness\\(mm)}} & {\makecell{No. of\\Patients}} & \makecell{Centralized\\training} & \makecell{Federated\\learning} & PPPML-HMI \\
 \hline
 \multirow{1}{*}{\makecell{1}} & A & 2 & 392 & 0.97 & / & /\\
 \hline
 \multirow{2}{*}{\makecell{2}} & A & 2 & 196 & \multirow{2}{*}{0.95} & \multirow{2}{*}{0.68} & \multirow{2}{*}{0.94}\\
 \multirow{2}{*}{} & B & 5 & 196 & & \\
 \hline
 \multirow{3}{*}{\makecell{3}} & A & 2 & 174 & \multirow{3}{*}{0.92} & \multirow{3}{*}{0.77} & \multirow{3}{*}{0.89}\\
 \multirow{3}{*}{} & B & 5 & 68 & & \\
 \multirow{3}{*}{} & C & 10 & 150 & &\\
 \hline
\end{tabular}
\end{table}

\begin{table*}[!htb]
    \caption{Description of the COVID-19 dataset. There are 5 hospitals (A$\sim$E) and each of them owns data generated by different CT scanners and settings.}
    \scriptsize
    \label{table1}
    \centering
    \begin{tabular}{c c c c c c c}
 \hline
 User ID & System label & Manufacturer & Model & Slice thickness (mm) & No. of Patients & No. of Total\\
 \hline
 \multirow{3}{*}{\makecell{A}} & P\_B16\_2.0 & Philips & Brilliance 16 & 2.00 & 5 & \multirow{3}{*}{\makecell{12}}\\
 \multirow{3}{*}{} & P\_B16\_3.3 & Philips & Brilliance 16 & 3.30 & 1 & \multirow{3}{*}{}\\
 \multirow{3}{*}{} & P\_B16\_7.5 & Philips & Brilliance 16 & 7.50 & 6 & \multirow{3}{*}{}\\
 \hline
 \multirow{2}{*}{\makecell{B}} & P\_I256\_1.0 & Philips & iCT 256 & 1.00 & 114 & \multirow{2}{*}{\makecell{119}}\\
 \multirow{2}{*}{} & P\_I256\_5.0 & Philips & iCT 256 & 5.00 & 5 & \multirow{2}{*}{}\\
 \hline
 C & P\_I\_1.0 & Philips & Ingenuity CT & 1.00 & 9 & 9\\
 \hline
 \multirow{5}{*}{\makecell{D}} & G\_B\_1.25 & GE medical systems & BrightSpeed & 1.25 & 1 & \multirow{5}{*}{24}\\
 \multirow{5}{*}{} & G\_B\_5.0 & GE medical systems & BrightSpeed & 5.00 & 4 & \multirow{5}{*}{}\\
 \multirow{5}{*}{} & G\_CT520\_1.25 & GE medical systems & Optima CT520 & 1.25 & 11 & \multirow{5}{*}{}\\
 \multirow{5}{*}{} & G\_CT540\_1.25 & GE medical systems & Optima CT540 & 1.25 & 7 & \multirow{5}{*}{}\\
 \multirow{5}{*}{} & G\_CT750\_5.0 & GE medical systems & Discovery CT750 HD & 5.00 & 1 & \multirow{5}{*}{}\\
 \hline
 \multirow{2}{*}{\makecell{E}} & S\_SDE\_1.0 & SIEMENS & SOMATOM Definition Edge & 1.00 & 10 & \multirow{2}{*}{\makecell{16}}\\
 \multirow{2}{*}{} & S\_SS\_2.0 & SIEMENS & SOMATOM Scope & 2.00 & 6 & \multirow{2}{*}{}\\
 \hline
\end{tabular}
\end{table*}

\subsection{PPPML-HMI is generalizable with various numbers of users and samples on the classification task.}

To show the robustness of PPPML-HMI when varying the number of users and the sample sizes, we simulated three sets of data partitions on the RAD-ChestCT dataset \cite{draelos2021machine, draelos_rachel_lea_2020_6406114} according to the slice thickness of CT scans and labeled them as Split 1, Split 2, and Split 3 as shown in $\bold{Table} \ref{table_rad}$. Split 1 had only one user with 392 CT scans, which represented the centralized training scenario. Split 2 had two users with an equal number of CT scans (196 and 196) but with different slice thicknesses (2 mm and 5 mm). Split 3 had three users with a varied number of CT scans (174, 68, and 150) and slice thicknesses (2 mm, 5 mm, and 10 mm) simultaneously. For each split, the goal was to train models for classifying healthy people and patients based on CT scans. Compared to centralized training, FL led to a drastic reduction in the averaged accuracy at Split 2 and Split 3, where PPPML-HMI showed less reduction and better performance compared to FL. These results provided evidence of the robustness of PPPML-HMI when varying the number of users and the sample sizes.

\subsection{PPPML-HMI achieved personalization for federated heterogeneous segmentation of lung infections by COVID-19}\label{secCC}

\begin{figure*}[htb]
\centering
\includegraphics[width=0.8\textwidth]{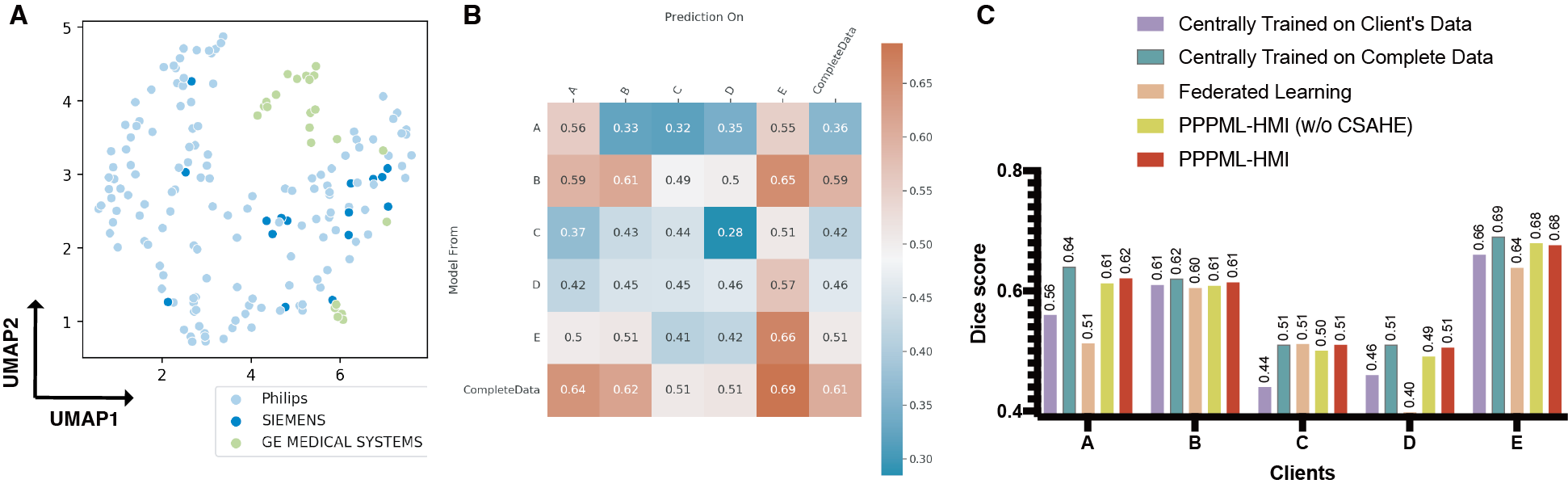}
\caption{A) Dimension reduction and clustering with UMAP according to the manufacturer indicated that CT scans generated by different CT scanners had significant inherent differences. B) Heatmap showed the Dice score of segmentation when applying models trained centrally on the data of each hospital. C) Barplot showed the Dice score of models trained centrally, with federated learning, and with PPPML-HMI.}
\label{fig3}
\end{figure*}

Since the classification task was relatively easy, we further applied PPPML-HMI to a real-world case, namely the segmentation of lung infections by COVID-19. There were five hospitals in the real-world case, labeled as $A\sim E$, each of which used CT scanners of different models (Brilliance 16, iCT 256, Ingenuity CT, BrightSpeed, Optima CT520, Optima CT540, Discovery CT750 HD, SOMATOM Definition Edge, and SOMATOM Scope) from three manufacturers (Philips, GE medical systems, and SIEMENS). Hospitals also used different settings, such as slice thickness (1.00 nm, 1.25 nm, 2.00 nm, 3.30 nm, 5.00 nm, and 7.50 nm), and provided various numbers of data ranging from 9 to 119 as shown in $\bold{Table}\ \bold{\ref{table1}}$. Different CT scanners and settings used by five hospitals led to inherent differences in the generated CT scans. As shown in $\bold{Figure}\ \bold{\ref{fig3}A}$, we performed the dimension reduction and clustering on the original CT scan data provided by the hospital with the uniform manifold approximation and projection (UMAP) \cite{mcinnes2018umap} and assigned different colors to visualize the inherent differences between data according to the manufacturer. Based on the results, CT scans generated by the CT scanners manufactured by GE medical system showed significant differences from that of other manufacturers, including Philips and SIEMENS, and indicated that inherent differences existed in CT scans generated by different hospitals with diverse CT scanners. 

To show the effectiveness of PPPML-HMI on federated heterogeneous medical image analysis, we compared different approaches, including the centralized training using complete data from all hospitals, the independent training using data from each hospital respectively, and the most classical and famous FL algorithm namely FedAvg as shown in $\bold{Figure}\ \bold{\ref{fig3}}$. Without taking the data privacy issues into consideration, the centralized training worked best when we collected complete data from all users $A \sim E$ to train the server model and performed the prediction on each user's data ($A_{Dice}=0.64, B_{Dice}=0.62, C_{Dice}=0.51, D_{Dice}=0.51, E_{Dice}=0.69, $) as shown in $\bold{Figure}\ \bold{\ref{fig3}B}$. However, when the privacy issue matters, the server could not collect data from users, thus centralized training could not be performed. Moreover, since data exists in a distributed manner, the heterogeneous data might cause problems in the FL scenario. To train models with heterogeneous data in FL, as one solution, using each user's own data to train independent models resulted in a significant performance reduction ($\triangle_{Dice} = -0.08, -0.01, -0.07, -0.05, -0.03$ for users $A \sim E$ respectively). Meanwhile, transferring a model trained on one user's data to another user showed even worse performance and indicated poor generalization ability of models trained with such a method, e.g. applying the model trained on user $C$ to user $D$ only yielded a Dice score of $0.28$. These results provided further evidence of the strong heterogeneity in data across users in the real-world case.

Since training a model with each user's own data did not meet the need for clinical-grade performance, training a model using the information in all users' data but without sharing the raw data was necessary. With that, FL was the most intuitive solution. However, because of the strong heterogeneity of users' data, the server model trained by FL ($A_{Dice} = 0.51, D_{Dice} = 0.39, E_{Dice} = 0.63$) performed even worse on users $A$, $D$, and $E$ than the independently trained model using only the users' own data ($A_{Dice} = 0.56, D_{Dice} = 0.46, E_{Dice} = 0.66$) as shown in $\bold{Figure}\ \bold{\ref{fig3}C}$. Meanwhile, user $C$ had only 9 samples and the data heterogeneity was not as significant as those between other users' data ($\bold{Figure}\ \bold{\ref{fig3}A}$), thus the server model trained with FL could achieve similar performance as the centralized training with complete data only at user $C$ ($\bold{Figure}\ \bold{\ref{fig3}B,C}$). 

In contrast to FL, similar to MAML, the server model generated by PPPML-HMI was a good initialization, which could learn the common features in heterogeneous data and could be easily adapted to local user's data by a few local training steps. As shown in $\bold{Figure}\ \bold{\ref{fig3}C}$, with the server model generated by PPPML-HMI as an initialization, only $\gamma$ ($\gamma<5$) steps of local training on the user's data could adapt the server model to local user's data ($A_{Dice} = 0.62, B_{Dice} = 0.61, C_{Dice} = 0.51, D_{Dice} = 0.51, E_{Dice} = 0.68$) and achieve a similar performance as the centralized training with complete data and better performance than FL under the same total number of epochs. Additionally, the improvement of PPPML-HMI compared to FL was most significant for users $A$ and $D$, as the data of both users showed the most significant data heterogeneity from the data of other users as shown in $\bold{Figure}\ \bold{\ref{fig3}A}$. 

To further understand the differences in performance between methods, we visualized the predicted segmentation of the lung infections on a high-quality sample (A000069) and a low-quality sample (A000075), respectively. The low-quality sample had significantly worse clarity and resolution of CT images than the high-quality sample as shown in $\bold{Figure}\ \bold{\ref{fig4}A}$. Orange, green, and yellow were used to represent true positives, false positives, and false negatives respectively compared to the ground truth. For A000069, the independently trained model on the associated user's data gave enormous false positives, while the model trained with FL predicted a large number of false negatives. In contrast, the personalized model from PPPML-HMI provided competitive performance as the centrally trained model with complete data. For A000075, the personalized model from PPPML-HMI predicted more true positives compared to the centrally trained model with the complete data and rescued more false negatives compared to the model trained with FL. In summary, PPPML-HMI allowed model personalization to each user with heterogeneous data and achieved competitive performance as the centralized training with complete data.

\subsection{PPPML-HMI protects privacy of CT scans}\label{secCC}

There are two types of attackers in our setting: 1) attackers who can intercept messages sent from any users to the server and between any users (type I), and 2) honest-but-curious (HBC) attackers that are part of the users of PPPML-HMI (type II) as in $\bold{Definition\ 1}$. 

$\bold{Definition\ 1}$. The HBC attacker is a legitimate participant in a communication protocol who will not deviate from the defined protocol but will attempt to learn all possible information from legitimately received messages \cite{paverd2014modelling}.

\begin{figure}[htb]%
\centering
\includegraphics[width=0.5\textwidth]{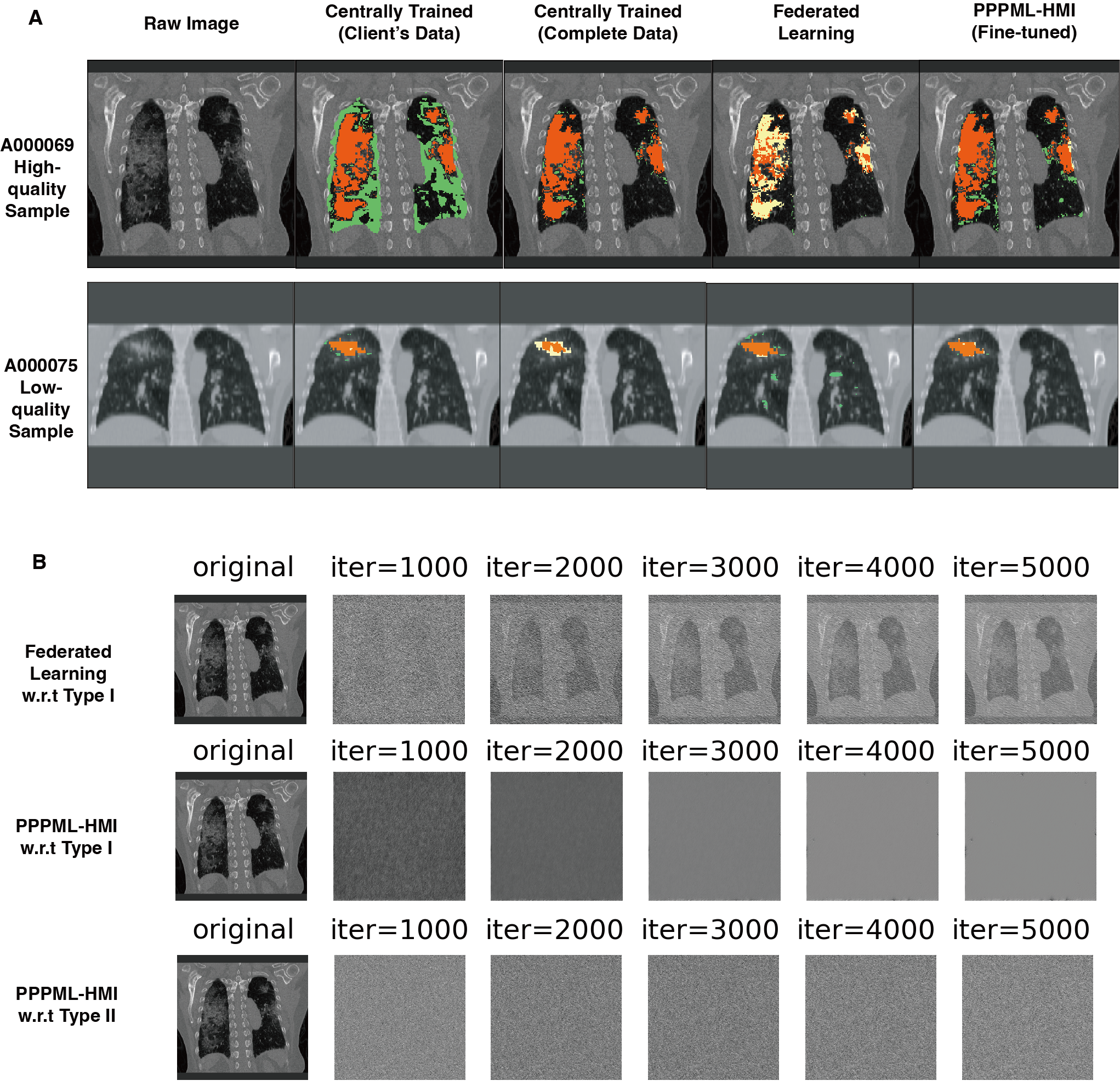}
\caption{A) Visualization of predicted segmentation mask on the high-quality sample A000069 and low-quality sample A000075 (Orange: true positives, Green: false positives, Yellow: false negatives). B) Visualization of the dummy data from the iDLG attack on FL and PPPML-HMI against both types of attackers.}
\label{fig4}
\end{figure}

Sensitive information could be deciphered from medical images, such as tissue patterns and lesions, which could compromise patients’ privacy \cite{kaissis2020secure}. CT scans of COVID-19 patients require even stronger privacy protection. To show that PPPML-HMI protected privacy from the CT scans of COVID-19 patients and resisted both types of attackers, we applied the iDLG ($\bold{Method}$) method to simulate an attacker reconstructing the training images by stealing the gradients transmitted between users and the server. During each attack, we initialized dummy data and used the gradient transmitted between the user and server to update the dummy data. We performed the iDLG attack for FL, where type I attackers exist and PPPML-HMI, where both types of attackers exist, respectively. Then, we visualized the dummy data every 1000 iterations, as shown in $\bold{Figure}\ \bold{\ref{fig4}B}$.

In practice, since the server is usually controlled by a third party, the users could not completely trust the server. As shown in $\bold{Figure}\ \bold{\ref{fig2}B}$, in FL, type I attackers could intercept the gradients sent between users and the server to learn the sensitive information of the corresponding users. With the CSAHE, only the initiator could communicate with the server and send the securely aggregated gradient. Though type I attackers could intercept the gradient between the initiator and the server, they only get the averaged gradient over all users. When type I attackers incept the information transmitted between users in CSAHE, they only get the ciphertext instead of the plaintext as all gradients transmitted in the loop are encrypted with HE. Hence, type I attackers could be resisted by PPPML-HMI. 

Type II attackers may exist in PPPML-HMI as gradients are transmitted between users as shown in $\bold{Figure}\ \bold{\ref{fig2}B}$. For example, an HBC user in the loop may try to recover sensitive information using the gradient transmitted from the previous user due to curiosity. As the key for HE is shared by all users in the loop, the HBC user could decrypt the gradient from the previous user and see the plaintext. However, the plaintext deciphered by any non-initiator users is always protected by the Gaussian noise added by the initiator, thus type II attackers could be resisted as shown in $\bold{Figure}\ \bold{\ref{fig4}B}$.

To demonstrate the results of gradient inversion, we applied iDLG to the associated segmentation model. As shown in $\bold{Figure}\ \bold{\ref{fig4}B}$, in terms of visual results, the iDLG attack on FL could effectively reconstruct the CT images in the training data, while the iDLG attack on PPPML-HMI was effectively blocked. Overall, PPPML-HMI protected privacy by blocking the reconstruction of sensitive medical images.

\section{Discussion}
In this paper, we present a novel, robust and open-source method for personalized and privacy-preserving federated heterogeneous medical imaging analysis. PPPML-HMI is a training paradigm similar to FL, which has no task-specific requirements and does not require any modifications to the existing DL models. With the nature of open-source, users of PPPML-HMI only need to apply PPPML-HMI as a plug-in to their neural network models as how they work with FL to achieve personalized and privacy-preserving federated learning. Based on the results of the simulated classification task on the RAD-ChestCT dataset and the real-word segmentation task based on the COVID-19 dataset, we believe that PPPML-HMI could be applied to any potential medical imaging problem with different DL methods, especially for those with heterogeneous data and the need for federation and privacy protection, as the segmentation method we adopted in the real-world case was also a general method for segmentation of lung, tracheal, vascular and so on. 

Though we applied PPPML-HMI in both simulated and real-world COVID-19 cases, all experiments were conducted in a laboratory environment, meaning all practical conditions were in the ideal state, including the computing power of the users' machines and the communication consumption between machines. As shown in $\bold{Table}\ref{table3}$, PPPML-HMI showed slightly higher requirements for the training time and similar memory and GPU compared to FL, meaning that PPPML-HMI could still work in the case that FL works. Meanwhile, integrating personalization and HE-based privacy protection in PPPML-HMI brought an additional 35.5\% computation time compared to FL, thus finding new solutions to accelerate could be one of the main research directions in the future.

\begin{table}[!htb]
    \caption{Comparison of computation resource requirements on the COVID-19 task. $n$ is the number of users.}
    \label{table3}
    \scriptsize
    \centering
    \begin{tabular}{ccccc}
 \hline
 Method & \makecell{Training\\Time (hours)} & \makecell{Virtual\\Memory (GB)} & \makecell{Physical\\Memory (GB)} & GPU (GB)\\
 \hline
 \makecell{Centralized\\training} & 320.21 & 70.26 & 3.34 & 26.80\\
 \hline
 \makecell{Federated\\learning\\($n=5$)} & 110.37 & 54.91 & 3.36 & 26.80 \\
 \hline
 \makecell{PPPML-HMI\\($n=5$)} & 149.55 & 54.97 & 3.43 & 26.80 \\
 \hline
\end{tabular}
\end{table}

Due to the special design of PPPML-HMI, it can only be applied when the number of clients $\ge 3$, meaning that PPPML-HMI is vulnerable when the number of clients equals 2. Suppose that the initiator is the type II attacker and there are two clients, then the initiator could decode the exact gradient of the other client and thus do the gradient inversion attack as shown in the classic FL setting. Though PPPML-HMI is not exactly private for the case where there are only two clients, it could work as designed when the number of clients $\ge 3$

Regardless of the issue in computing resources and vulnerability, we need to take more practical problems into consideration, such as the imbalance of computing power among hospitals, the deviation of data quality of different hospitals, the latency in network communication between hospitals and the server, and so on. Those practical problems have not been addressed in this work as we were focusing on a learning paradigm. Still, we will solve those realistic obstacles and further improve PPPML-HMI in future work.

{
\bibliographystyle{IEEEtran}
\bibliography{reg}
}

\end{document}